\documentclass[a4paper,12pt]{iopart}

\usepackage{graphicx,amssymb,color} 
\usepackage[utf8]{inputenc}
\usepackage[english]{babel}

\newcommand{\Notfall}{\hspace*{-2.2cm}}
\newcommand{\Halbnotfall}{\hspace*{-1.1cm}}

\newcommand{\hb}{\hat{b}}
\newcommand{\hH}{\hat{H}}

\newcommand{\ko}[2]{[#1, #2]}
\newcommand{\ako}[2]{\{#1, #2\}}

\begin{document}

\title[]{Transport induced melting of crystals of Rydberg dressed atoms in a one dimensional lattice}
\author{Achim~Lauer$^1$, Dominik~Muth$^1$ and Michael~Fleischhauer$^1$}
\address{$^1$ Fachbereich Physik und Forschungszentrum OPTIMAS, Technische Universit\"at Kaiserslautern, D-67663 Kaiserslautern, Germany}
\ead{mfleisch@physik.uni-kl.de}
 
\begin{abstract}
We discuss the many-body physics of an ensemble of  Rydberg dressed atoms with van der Waals dipole-dipole interactions in a one-dimensional
lattice. Using a strong coupling expansion and numerical density-matrix renormalisation group simulations, we calculate the many-body phase diagram. A devil's staircase structure emerges with Mott-insulating phases at any rational filling fraction. Closed analytic expressions are given for the phase boundaries in second order of the tunnelling amplitude and shown to agree very well with the numerical results.
The transition point where the incompressible phases melt due to the kinetic energy term depends strongly on the denominator of the filling fraction and varies over many orders of magnitude between different phases. 
\end{abstract}

\pacs{  37.10.Jk, 
	32.80.Ee, 
	64.70.Tg, 
	05.30.Jp 
}
 
\date{\today}
 
\maketitle

\section{Introduction}

Recently many-body systems with non-local, power-law interactions gained
considerable interest  as the non-local coupling can give rise 
to quantum phases that do not exist for point-like interactions \cite{Lahaye2009}.
 For example in one spatial dimension
repulsive dipole-dipole or van der Waals interactions of atoms can lead to a variety of crystalline ground state phases with less than unity filling 
in the presence of a commensurable periodic lattice \cite{Batrouni2006,Pupillo2010,Schachenmayer2010,Lesanovsky2011}.  For very 
strong lattice confinements the filling fraction forms
a so-called complete devils staircase as function of the chemical potential $\mu$, i.e. every rational filling between $0$ and $1$ is stable for a finite interval of $\mu$ and these intervals form a dense staircase \cite{Burnell2009}.
Power-law interactions arise e.g. for dipolar atoms or molecules \cite{Lahaye2009}. A 
very interesting alternative approach are Rydberg gases for which there has been considerable experimental progress in the recent years 
\cite{Tong2004, Vogt2007, Raitzsch2008, Johnson2008, Reetz-Lamour2008, Urban2009, Gaetan2009}. While most previous experiments implemented schemes where
Rydberg excitations were created by continuous near-resonant laser driving, alternative proposals have been put forward to use far-detuned light fields to admix
a small component of a Rydberg state to the ground state of atoms 
\cite{Pupillo2010,Johnson2010,Henkel2010}. The potential advantage of this ``Rydberg-dressing'' is that the interaction strength can be tuned to a certain extend and is reduced to an energy scale where the center-of-mass motion of atoms can become relevant. Furthermore continuous driving by a near-resonant
laser does in general not conserve the number of Rydberg excitations, while in the case of Rydberg-dressing the number of atoms are to good approximation a conserved quantity. While the non-local repulsive interaction
favors crystalline structures with non-unitary filling in lattice gases,
hopping between lattice sites induced by tunnelling of the atoms can lead to a melting of the crystals \cite{Weimer2010,Sela2011}.

In the present paper we analyse the melting process induced
by hopping of Rydberg-dressed atoms. In particular we derive the
phase boundaries of stable crystalline structures as a function of the hopping amplitude $J$ 
both analytically and with exact numerical simulations. 
To this end we employ a second order strong coupling expansion on the one hand and numerical density-matrix renormalisation group (DMRG) simulations, adapted to long-range
interactions, on the other.  Both show excellent agreement up to second order in $J$.
A corresponding strong-coupling analysis of the phase diagram has been discussed before for the case of dipolar $1/r^{3}$ interactions
in \cite{Burnell2009}, however no explicit expressions were given. We here derive
analytic expressions which in the case of fast decaying potentials, such as the $1/r^6$ van der Waals potential, can in very good
approximation be written in a compact closed form. Furthermore we provide for the first time a comparison to numerical data 
from DMRG simulations adapted to long-range interactions.

As a starting point we consider the generalised Bose-Hubbard model in one dimension
\begin{equation}
 \hH =  \underbrace{\frac{1}{2}\sum_{i\ne j} V(|r_i-r_j|)\hb^\dagger_i\hb_i\hb^\dagger_j\hb_j}_{\hH_0} - \underbrace{ J \sum_{i} \big(\hb^\dagger_i\hb_{i+1} + \hb^\dagger_{i+1}\hb_i\big)}_{\hH_J}.
\label{eq:ham}
\end{equation}
For convenience we have set $\hbar=a=1$, where $a$ is the lattice constant. $\hb_i,\hb_i^\dagger$ are the annihilation and creation operators at site $i$, and
$J$ is the hopping amplitude between neighbouring lattice sites describing the tunnelling of atoms between adjacent
lattice sites \cite{Jaksch1998}. 
Most parts of our discussion are valid for any convex interaction potential, i.e
\begin{equation}
 V(r+1)+V(r-1)\ge 2 V(r).
\label{eq:convex}
\end{equation}
However to be specific we discuss in the following power-law potentials of the form
\begin{equation}
 V(r)= \frac{\tilde{C}_\beta}{|r|^\beta}
\label{eq:potential}
\end{equation}
with $\beta\in\mathbb{N}\backslash \{1\}$ and $\tilde{C}_\beta>0$ being the interaction coefficients. 
$\beta=6$ corresponds to a van der Waals type 
interaction which results from virtual (i.e. off-resonant)  long-wavelength electromagnetic transitions in the manifold of Rydberg states. If there is an accidental or engineered 
resonance, a so called F{\"{o}}rster resonance, also the case $\beta=3$ can be of relevance. The model interaction potential (\ref{eq:potential})  diverges for $r\to 0$ and thus two particles cannot sit on the same lattice site. To very good approximation the same is 
true if one considers the exact interaction potential between Rydberg atoms which deviates from the above power law for small distances (see e.g.\cite{Schwettmann2006,Stanojevic2006}). 
In fact for the Rydberg-dressing scheme one finds an effective two-body interaction potential between atoms at positions $\mathbf{r}_i$ and $\mathbf{r}_k$ \cite{Pupillo2010,Johnson2010,Henkel2010}
\begin{equation}
 V(\mathbf{r}_i,\mathbf{r}_k) = \frac{\eta^2 C_6}{|\mathbf{r}_i-\mathbf{r}_k|^6 + R_c^6}
=\frac{{\tilde C}_6}{r^6 + R_c^6},\quad\qquad r=|\mathbf{r}_i-\mathbf{r}_k|.
\end{equation}
Here $C_6$ is the Rydberg interaction coefficient and $R_c=\left(C_6/(2\Delta)\right)^{1/6}$ describes a characteristic length scale of the interaction potential,
where $\Delta\gg|\Omega|$ is an effective detuning of the off-resonant laser driving. $\eta\sim \Omega^2/\Delta^2 \ll 1$ accounts for the small admixture of the Rydberg state.
Considering e.g. $^{87}$Rb with a Rydberg state of principle quantum number around $n=60$ and detunings of a few tens of MHz one finds a value of $R_c$ on the order of
one to a few $\mu$m. In a recent experiment atoms were loaded into optical lattices and excited to Rydberg states with $n=55-80$ \cite{Viteau2011}. In this experiment the lattice spacing was tunable 
between $0.5 \mu$m and $13 \mu$m. Thus the cut-off length $R_c$ may become smaller than the average distance between particles given by the lattice constant devided by the average occupation number per lattice
site. In this case its effect can be disregarded and this is the parameter regime we are considering here.
For that reason $\hb_i,\hb_i^\dagger$ 
are assumed to describe hard-core-bosons with commutation relations
\begin{eqnarray}
 \ko{ \hb_i}{\hb_j^\dagger}&=\ko{ \hb_i}{\hb_j}&=\ko{ \hb_i^\dagger}{\hb_j^\dagger} = 0 \quad i\ne j,\\
 \ako{\hb_i}{\hb_i}&=\ako{\hb_i^\dagger}{\hb_i^\dagger} &= 0,\\
 \ako{\hb_i}{\hb_i^\dagger}&=1.
\end{eqnarray}
We note that although we discuss here the particular case of the integer value $\beta=6$,
the qualitative structure of the phase diagram holds true for any positive value of $\beta$.

\section{Ground state for $J=0$ -- classical limit}
a.marueg@web.de
Let us first consider the ground state of Hamiltonian (\ref{eq:ham}) in the limit of vanishing hopping, i.e. $J\to 0$.
In this limit the Hamiltonian is diagonal in the number basis, where each lattice site contains exactly zero or exactly one particle, which corresponds to
a classical situation. Consequently the energy can be minimised just by finding the configuration of classical particles that gives the smallest interaction energy. 
As the interaction potential is convex (see equation (\ref{eq:convex})), the minimum energy for commensurable filling fractions $q=\frac{m}{n}\le1$, where $m,n\in \mathbb{N}$ are relatively prime, is attained by a regular pattern with unit cells of size $n$
 \cite{Hubbard1978}. For a given chemical potential the ground state is such a phase with rational $q$. For these values of $q$ the corresponding phase is incompressible, i.e., particle as well as hole excitations require finite energy, i.e. $\mu^+>\mu^-$, where
\begin{equation}
 \mu^\pm(q)=\left.\frac{\partial E}{\partial N}\right\vert_{\rho=q\pm0} = \pm\left(E^\pm(q)-E(q)\right).
\label{eq:mudef}
\end{equation}
Here $E(q)$ is the ground state energy for filling fraction $q$. $E^\pm(q)$ is the corresponding energy where one particle has been added ($E^+(q)$) or respectively removed ($E^-(q)$). $\rho$ is the average number of particles per lattice site. Because of particle-hole symmetry,
\begin{equation}
 \mu(q) = \Omega-\mu(1-q),
\end{equation}
where $\Omega$ is the energy per particle in the completely filled lattice ($\Omega$ is the Riemann zeta function $\zeta(\beta)$ in the case of \eref{eq:potential}), 
it is sufficient to consider only filling fractions $q\le \frac{1}{2}$.

\begin{figure}[tb!]
\centering
 \includegraphics[scale=0.5]{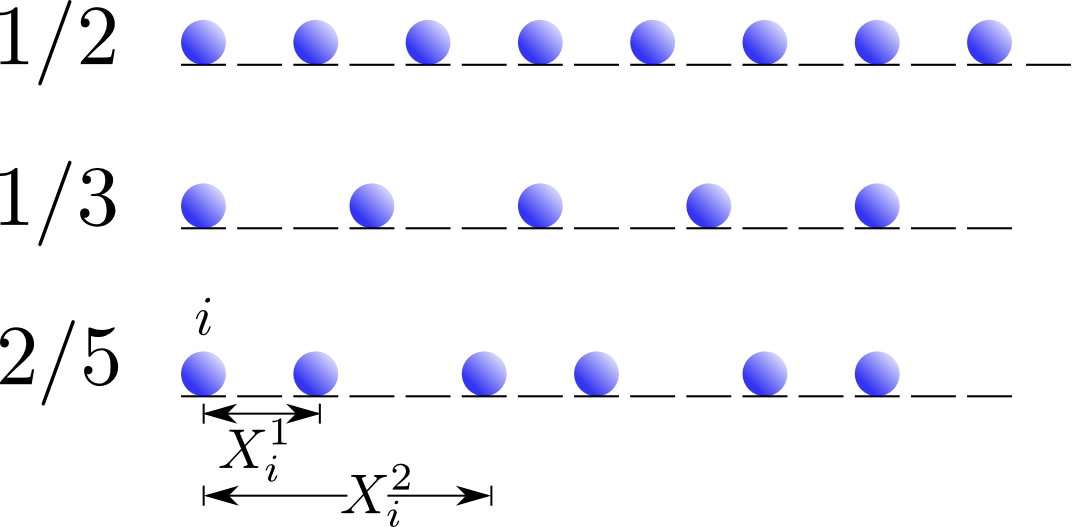}
\caption{Ground states for filling fractions $q=\frac{1}{2},q=\frac{1}{3}$ and $q=\frac{2}{5}$ at $J=0$.
\label{fig:groundstates}}
\end{figure}

The phase-diagram in zeroth order of the hopping $J$ has been discussed some time ago by Bak and Bruinsma \cite{Bak1982}. Here, we will briefly review their 
results and we will make use of their notation: $X_i^0$ denotes the position of the $i$-th particle. $X_i^p=X_p^0-X_i^0$ describes the distance between the $p$-th and the $i$-th particle where the particles are numbered from the left to the right. In the ground state of $\hH_0$ 
Hubbard's solution \cite{Hubbard1978} shows that all possible distances are given by
\begin{equation}
 X_i^p = r_p \quad {\rm or}\quad X_i^p = r_p +1,
\end{equation}
where $r_p < \frac{p}{q} < r_p + 1$. If $ \frac{p}{q} \in \mathbb{N}$ then $X_i^p = r_p = \frac{p}{q}$ and all particles are equivalent. Examples of ground states are shown in \fref{fig:groundstates}. Due to the convexity of the interaction, all nearest-neighbour separations are maximally close to the average separation, i.e. they are either $\lfloor\frac1q\rfloor$ or $\lceil\frac1q\rceil$.

\begin{figure}[t!]
\centering
a)\quad\includegraphics[width=.4\textwidth]{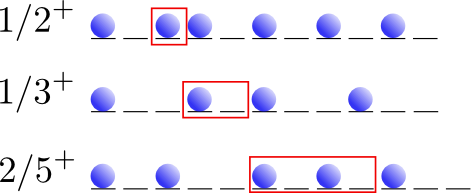}\quad
b)\quad\includegraphics[width=.4\textwidth]{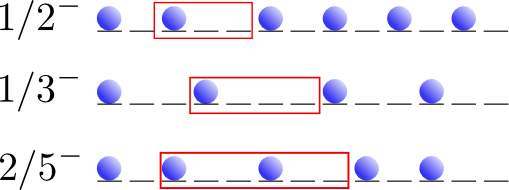}
\caption{Defects for $q=\frac{1}{2},q=\frac{1}{3}$ and $q=\frac{2}{5}$. The boxes mark the extension of the defect. Outside of the boxes the configuration of particles is exactly the same as in the commensurable case, see \fref{fig:groundstates}. 
\label{fig:hopping-solitons_zero}}
\end{figure}

\begin{figure}[t!]
\centering
a)\quad\includegraphics[width=.43\textwidth]{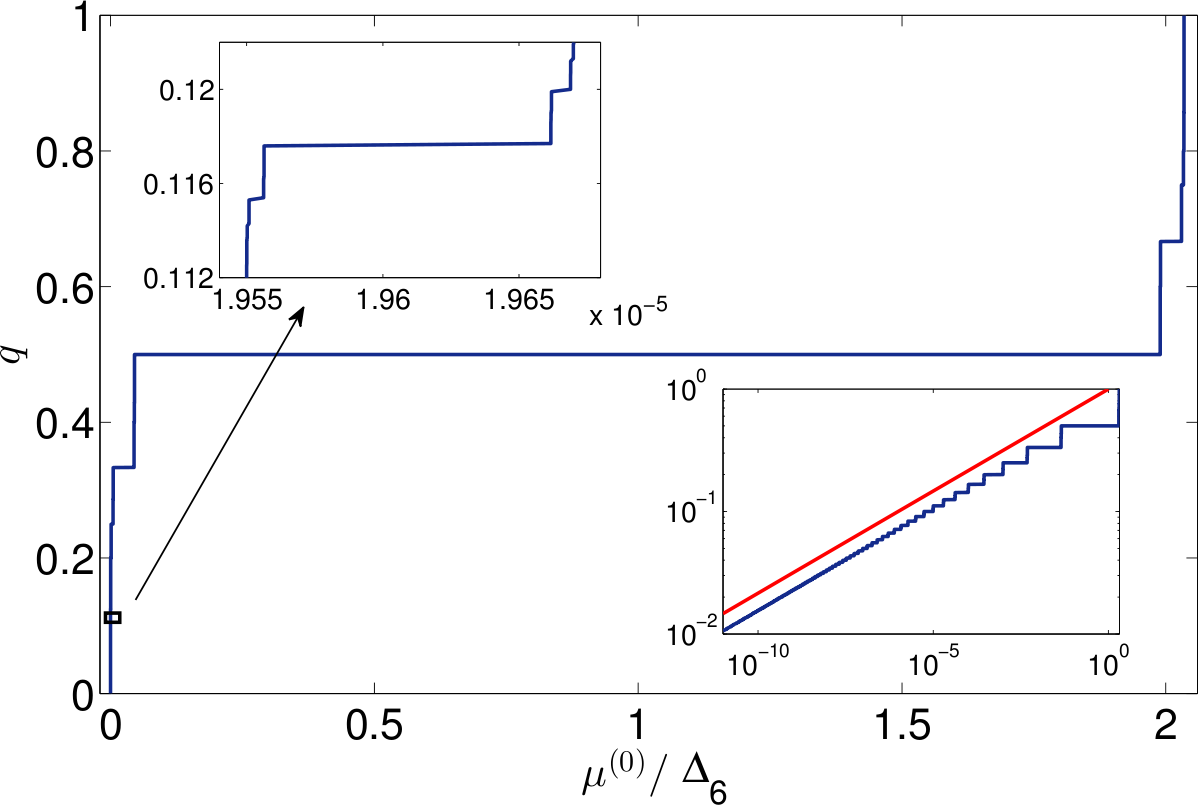}\quad
b)\quad\includegraphics[width=.43\textwidth]{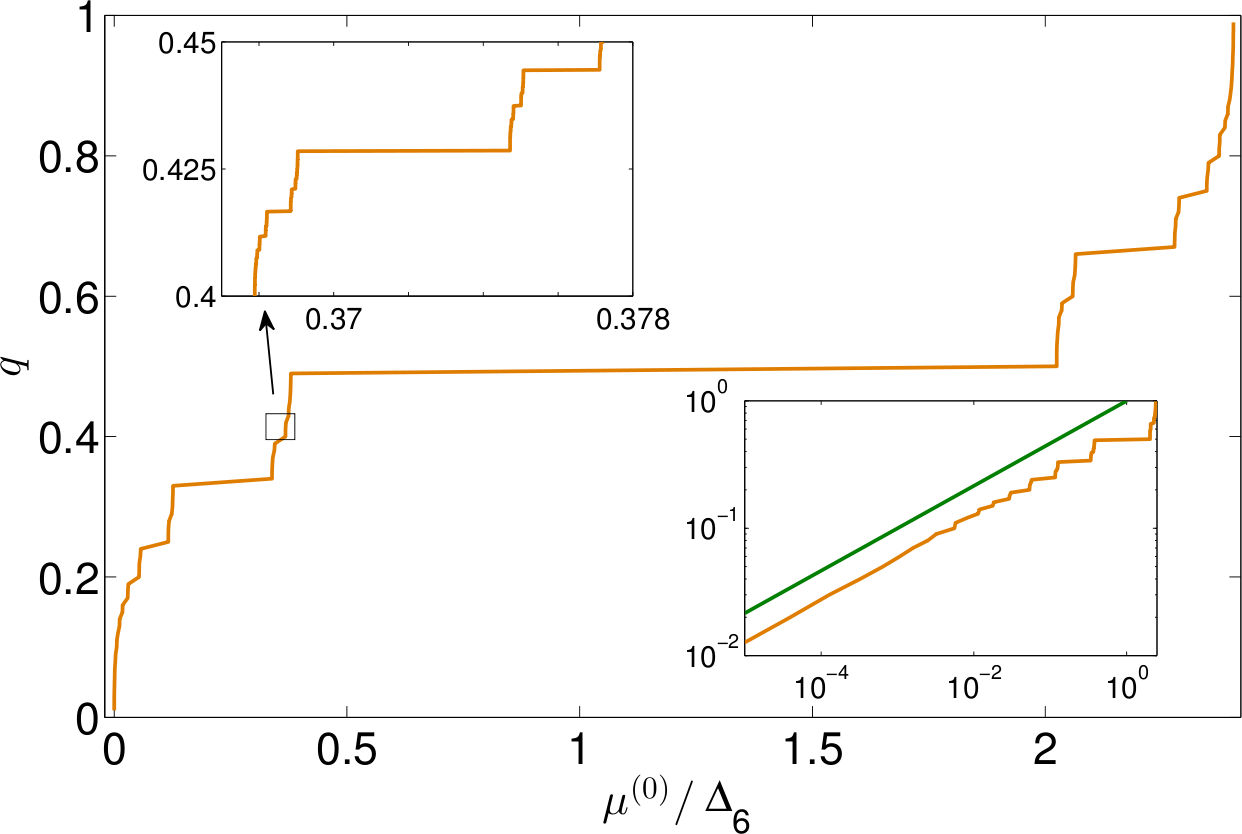}
\caption{Devil's staircase: The filling fraction $q$ in the ground state $J=0$ is plotted versus the chemical potential $\mu^{(0)}$ for a van der Waals interaction potential $\beta=6$ (a) and a F{\"{o}}rster resonance $\beta=3$ (b) in units of the level shift $\Delta_\beta \equiv \tilde{C}_\beta/a^\beta$. The insets to the upper left emphasise the repeating features of the devil's staircase. The insets at the right bottom show this equation of state in a double logarithmic plot. The red line in (a) and the green line in (b) correspond to $q=(\mu^{(0)}/\tilde C_\beta)^{-\beta}$.
\label{fig:devil_staircase}}
\end{figure}

The chemical potential for $J=0$ is given by
\begin{equation}
 \mu^{(0)}_\pm(q)= \pm \left( \langle q^\pm| \hH_0 | q^\pm\rangle - \langle q| \hH_0 | q\rangle \right).
\label{eq:mu0gen}
\end{equation}
$|q\rangle$ denotes the ground state of $\hH_0$ for filling fractions $q$, whilst $|q^\pm\rangle$ denotes the corresponding ground states where one particle has been added $|q^+\rangle$, respectively removed $|q^-\rangle$. Although both expectation values in \eref{eq:mu0gen} are infinite in the thermodynamic limit, their difference is finite, and can be calculated by summing the change in interaction energy particle by particle. The same is true for the first and second order corrections in the hopping amplitude $J$ discussed below. For filling fractions $q=\frac{m}{n}$ there will be $n$ defects for $|q^\pm\rangle$, because it is energetically favourable to break up an extra particle (or a hole) into $n$ defects each with fractional charge $\pm\frac1n$. Some of these defects are displayed in \fref{fig:hopping-solitons_zero}. In the thermodynamical limit they will be separated by arbitrarily large distances so that we can assume without approximation that they do not interact with each other at all. Note that $| q^\pm\rangle$ is not uniquely defined, but the position of defects is arbitrary, as long as they are well separated.
For filling fractions of the form $q=\frac{1}{n}$ the chemical potential is given by \cite{Bak1982}
\begin{equation}
 \mu_\pm^{(0)} = \pm\sum_{p=1}^\infty \Bigl( r_p\ V(r_p\mp1)- (r_p\mp1) V(r_p)\Bigr).
\label{eq:mu0}
\end{equation}
We can go a step further and evaluate \eref{eq:mu0} for the power-law potential \eref{eq:potential} analytically,
\begin{equation}
 \Halbnotfall\mu_\pm^{(0)}/\tilde{C}_\beta = 
\frac{(-1)^\beta \Psi^{(\beta-1)}(\frac{n\mp1}{n})}{n^\beta(\beta-1)! } + \frac{\zeta(\beta)}{n^\beta}
\pm \left(\frac{(-1)^{(\beta-1)} \Psi^{(\beta-2)}(\frac{n\mp1}{n})}{n^{(\beta-1)}(\beta-2)! } - \frac{\zeta(\beta-1)}{n^{(\beta-1)}}\right)
\label{eq:mu0power}
\end{equation}
using the digamma function $\Psi(z)= \frac{\Gamma'(z)}{\Gamma(z)}$ and its $l$-th derivative $\Psi^{(l)}(z)$. To describe more general filling fractions $q=\frac{m}{n}$ with $m\ne1$ as well, we only have to replace $r_p\rightarrow r_p+1$ for $\mu_+^{(0)}$ in \eref{eq:mu0} if $r_p\not\in \mathbb{N}$. Plotting the filling fraction as a function of the chemical potential gives rise to a complete devil's staircase (see \fref{fig:devil_staircase}). Every rational filling fraction between 0 and 1 has a finite stable range with respect to the chemical potential and the functional dependence $q(\mu)$ consists of a dense set of steps between these stable plateaus, a so-called devils staircase. The union of these intervals covers the full range of $\mu$. It is clearly seen that filling fractions with small denominator $n$ are the most stable as can be seen directly from equation \eref{eq:mu0power}. These intervals of stability depend crucially on the exponent $\beta$: for large $\beta$, the interval of the stable phase $q=\frac12$ overgrows by far the size of all the other phases. Note that for $n$ big enough $(n\mp1)/n$ is almost constant and hence the mean chemical potential $\mu^{(0)}\equiv
(\mu_+^{(0)}-\mu_-^{(0)})/2$ scales as $n^{-\beta}$. Thus below half filling the 
fraction $q$ scales as $(\mu^{(0)}/\tilde C_\beta )^{-\beta}$ as indicated in the inset of \fref{fig:devil_staircase}.

As mentioned in the introduction the true interaction potential between Rydberg-dressed
atoms becomes flat below a certain distance. As a consequence the atoms can be treated as hard-core bosons
only for sufficiently small energies, i.e. sufficiently small chemical potentials. The finite cut-off $R_c$ will lead
to a qualitative change of the phase diagram in parameter regions where the separation between atoms
becomes smaller than $R_c$. This is however only the case if the filling $q$ becomes large and
in particular exceeds the ratio $R_c/a$.

\section{Perturbation theory in $J$}

We are now interested in the melting of a crystal phase with increasing hopping rate $J$. Therefore we perform perturbation theory up to second order in $J$. Similar calculations have been done by Burnell et al. for dipolar interactions $V \sim 1/r^3$ \cite{Burnell2009}, but no compact analytic expression was given.

\begin{figure}[htb!]
\centering
a)\quad\includegraphics[width=.4\textwidth]{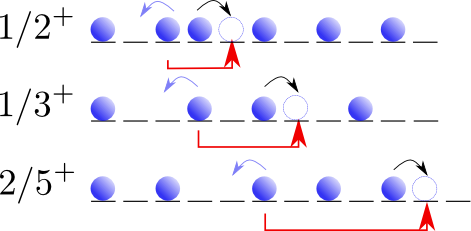}\quad
b)\quad\includegraphics[width=.4\textwidth]{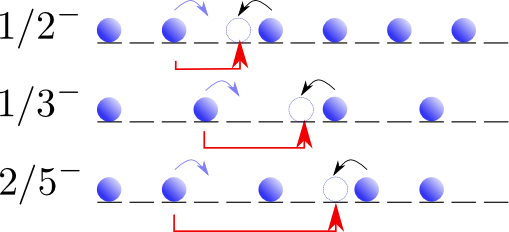}
\caption{Hopping of defects for $q=\frac{1}{2},q=\frac{1}{3}$ and $q=\frac{2}{5}$. The black arrow indicates the hopping of one particle and the red one corresponds to the resulting hopping of the defect. The blue ones are the alternative hopping-possibility, which would result in a hopping of the defect in the opposite direction.
\label{fig:hopping-solitons_one}}
\end{figure}

Let us start with first order processes in $J$. Then, the chemical potential reads $ \mu^{(1)}_\pm(q)= \pm \left( \langle q^\pm| \hH_J | q^\pm\rangle - \langle q| \hH_J | q\rangle \right)$. Hopping of any single particle in state $|q\rangle$  will not contribute to any energy correction, because the resulting state has no overlap with the $J=0$ ground state. The same is true for hopping of any but the $2n$ particles of the state $|q^\pm\rangle$  which sit at the left and right borders of any of the $n$ defects. Hopping of these particles will lead to a hopping of the respective defect by $n$-sites, see \fref{fig:hopping-solitons_one}. As the states with localised defects are degenerated with respect to $\hH_0$, we have to apply degenerate perturbation theory. Therefore, we look for a basis of these states in which $\hH_J$ is diagonal. This is given by states where all the defects are delocalised over many lattice sites (yet well separated from each other) with some quasi-momentum $k$. The energy is minimised for a state with quasi-momentum zero. The resulting first order correction to the phase boundary thus takes the simple form
\begin{equation}
 \mu^{(1)}_\pm(q)= \mp 2 n J.
\end{equation}

For small but nonzero $J$ gaps open between the incompressible phases of rational filling, because the kinetic energy 
gained by delocalisation of defects favours a finite density of defects.
At the tip of the phase $J$ becomes large enough such that the creation of particle and hole defects becomes favourable even at 
commensurate filling fractions and the crystalline phase melts. This is completely analogous to the ordinary Hubbard model. However, for 
any arbitrarily small but finite value of $J$ almost all phases with rational filling become unstable, leaving only a finite number 
(those with the smallest denominator $n$) and destroying the devils staircase.a.marueg@web.de

\begin{figure}[t!]
\centering
a)\quad\includegraphics[width=.375\textwidth]{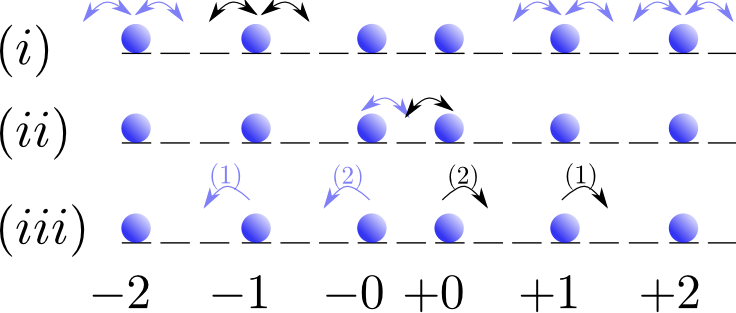}\quad
b)\quad\includegraphics[width=.425\textwidth]{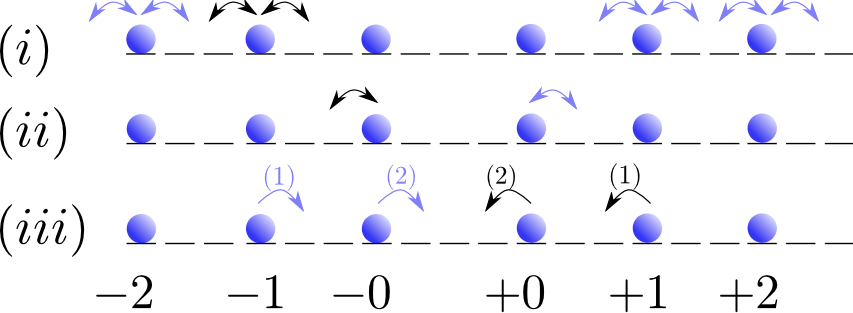}
\caption{All possible second order processes for $|q^+\rangle$ (a) and  $|q^-\rangle$ (b) at filling $q=\frac1n$, here $q=\frac13$. (i) generates an effective chemical potential for single particles that depends on the distance to the defect, i.e., the defect polarises the background. (ii) refers to a virtual deformation of the defect. (iii) corresponds to an effective hopping of the defects over two unit cells.
\label{fig:1-n_processes}}
\end{figure}

In second order perturbation the energy corrections are given by $ E^{(2)} = \sum_{\Phi\ne\Psi}\frac{\langle\Psi| \hH_J| \Phi\rangle\langle \Phi | \hH_J | \Psi\rangle}{E^{(0)}_\Psi-E^{(0)}_\Phi}$, where $|\Psi\rangle$ denotes the ground state of $\hH_0$ and $\{|\Phi\rangle\}$ is a complete set of orthonormal states. 
In the following we consider only filling fractions of the form $q=\frac{1}{n}$. In principle it is posa.marueg@web.desible to extend the calculations to other fractions. 
However, finding a general formula which describes all energy differences in the denominator for all hopping processes is a far more tricky task. 
In \fref{fig:1-n_processes} all possible processes for $q^\pm=\frac{1}{n}^\pm$ are shown. As has been seen in first order, the states with localised defects are degenerate with respect to $\hH_0$.
In degenerate second order perturbation theory the energy correction is given by $\langle q^\pm|\hH_J \hat P \hH_J|q^\pm\rangle/\big(E^{(0)}_{q^\pm}-E^{(0)}_\Phi\big)$, where $\hat P$ projects out the subspace of ground states of $\hH_0$. The state $|q^\pm\rangle$ has to be an eigenstate of $\hH_J$ restricted to the degenerate subspace. Therefore, intermediate states $|\Phi\rangle$ with $E^{(0)}_\Phi=E^{(0)}_{q^\pm}$ do not contribute to the energy corrections.
For states $|q\rangle$ only process (i) in \fref{fig:hopping-solitons_one} contribute. For
the commensurate state $|q\rangle$ one finds for the second order correction to the energy
\begin{equation}
 E^{(2)}(q) = 2J^2 N \frac{1}{\Delta E^{(0)}(q)}\label{eq:decomposition},
\end{equation}
where $N$ is the number of particles in the lattice. $\Delta E^{(0)}(q)=  E^{(0)}(q)-E_+^{(0)}(q)$ is the energy difference between the 
ground state $|q\rangle$ and the same state but with one particle hopped by one site. With the notation introduced above it can be evaluated to
\begin{eqnarray}
 \Delta E^{(0)}(q)  &=  \sum_{p=1}^{\infty}\Big(2 V(r_p)-V(r_p-1) - V(r_p +1)\Big)\\
&=\frac{2\zeta(\beta)}{n^\beta}-\frac{(-1)^\beta}{n^\beta(\beta-1)!}\! \left[\Psi^{(\beta-1)}\! \left(\frac{n-1}{n}\right)\! +\Psi^{(\beta-1)}\! \left(\frac{n+1}{n}\right)\right]\! .\nonumber
\end{eqnarray}
In order to obtain the (finite) chemical potential we have to calculate also the energy corrections $E^{(2)}(q^\pm)$ for the states with one extra particle or hole $|q^\pm\rangle$ in second order
perturbation theory. 
To this end let us have a closer look at all possible hopping processes displayed in \fref{fig:1-n_processes}. As has been done in first order, we will concentrate on a single defect multiplying the resulting contribution with its number $n$. Process $(ii)$ describes a virtual deformation of the defect while process $(iii)$ corresponds to an effective hopping of the defects over two unit cells. Both processes have only to be taken into account for one particle of the defect and one particle on its right side and therefore contribute only to a finite energy value. In the thermodynamical limit, process $(i)$ takes place for an infinite number of particles. For this reason, its contributions to the energy corrections will diverge. However, for a decaying potential the contribution to the energy correction of a particle far away from the defect will be the same as the contribution of a particle of the commensurate state $|q\rangle$. Subtracting now the energy corrections for states $|q\rangle$ and $|q^\pm\rangle$,  
these contributions of process $(i)$ will mostly cancel. We find
\begin{equation}
\Notfall  E^{(2)}(q^\pm) = 2 n J^2 \left\{\sum_{i=1}^{N_0^\pm} \left(\frac{1}{\Delta E^{(0)\pm}_{+i,-}} +\frac{1}{\Delta E^{(0)\pm}_{+i,+}}\right) +
\frac{1-\frac12(\delta_{n,2}\pm\delta_{n,2})}{\Delta E^{(0)\pm}_{+0,\mp}} +\cos(2 k) \frac{1}{\Delta E^{(0)\pm}_{+1,\pm}}\right\}
\label{eq:decomposition-pm},
\end{equation}
where $\Delta E^{(0)\pm}_{+j,-}$ denotes the energy difference between the ground state $|q^\pm\rangle$ and the same state, where the $j$-th particle right of the defect moves right (second subscript $+$) or left (second subscript $-$). For the meaning of position ``$+0$`` see \fref{fig:1-n_processes}.
We have assumed here that the defect is in the centre of the system and thus the summation limit is $N_0^\pm = 1/2[(N\pm1)q-2]$.
As $|q^\pm\rangle$ has to be an eigenstate of $\hH_J$ in the degenerate subspace the defects are completely delocalised. As in first order it turns out that the ground state has quasi-momentum $k=0$. Then, the second order correction to the chemical potential can be written as
\begin{eqnarray}
 \Notfall\mu^{(2)}_\pm = 
\pm2 n J^2 \Bigg\{
&\sum_{j=1}^{\infty}\Big[\frac{1}{\Delta E^{(0)\pm}_{+j,-}}+\frac{1}{\Delta E^{(0)\pm}_{+j,+}}-\frac{2}{\Delta E^{(0)}}\Big]-(1\mp\frac{1}{2n})\frac{2}{\Delta E^{(0)}}\label{eq:mu2_1}\\
&+\frac{1-\frac12(\delta_{n,2}\pm\delta_{n,2})}{\Delta E^{(0)\pm}_{+0,\mp}}\label{eq:mu2_2}\\
&+\frac{1}{\Delta E^{(0)\pm}_{+1,\pm} }\Bigg\},\label{eq:mu2_3}
\end{eqnarray}
Part \eref{eq:mu2_1} takes into account all processes of type $(i)$. Because of symmetry it is only necessary to sum over particles on the right side of the defect.  
The last term in this line is due to the fact that there are less particles in $|q^\pm\rangle$ taking part in process $(i)$ then there are particles in state $|q\rangle$. Part \eref{eq:mu2_2} describes process $(ii)$ and part \eref{eq:mu2_3} process $(iii)$. Calculating all the energy differences analogous to the case of $\Delta E^{(0)}$, the chemical potential  for filling fractions $q=\frac{1}{n}$ can finally be written in second order in $J$ as
\begin{eqnarray}
 \Halbnotfall\mu^{(2)}_{\pm} = \pm\, 2 n \frac{J^2}{\tilde{C}_\beta}\, \Bigg\{& \sum_{j=1}^\infty S_j^\pm(n)
 -\frac{2\mp\frac{1}{n}}{\frac{2\zeta(\beta)}{n^\beta}-\frac{(-1)^\beta}{ n^\beta(\beta-1)!}[\Psi^{(\beta-1)}(\frac{n-1}{n})+\Psi^{(\beta-1)}(\frac{n+1}{n})]}\nonumber\\
&+\frac{1-\frac12(\delta_{n,2}\pm\delta_{n,2})}{\frac{\zeta(\beta)}{n^\beta}+\frac{(-1)^\beta}{ n^\beta(\beta-1)!}[\Psi^{(\beta-1)}(\frac{n\mp1}{n})-\Psi^{(\beta-1)}(\frac{n\pm1}{n})-\Psi^{(\beta-1)}(\frac{n\mp2}{n})]}\nonumber\\
&+\frac{1}{\frac{2}{n^\beta}-\frac{1}{(n+1)^\beta}-\frac{1}{(n-1)^\beta}} \Bigg\},
\end{eqnarray}
where the terms $S_j^\pm(n)$ are given by
\begin{eqnarray}
 \Halbnotfall S_j^\pm(n) &=\Bigg\{\sum_{p=1}^j\left[\frac{2}{(pn)^\beta}-\frac{1}{(pn+1)^\beta}-\frac{1}{(pn-1)^\beta}\right]\Bigg\}^{-1}\nonumber\\
+&\Bigg\{ \sum_{p=1}^j\!\Big[\frac{1}{(pn)^\beta}\!-\!\frac{1}{(pn+1)^\beta}\!-\!\frac{1}{(pn-1)^\beta}\Big]\!+\!\frac{\zeta(\beta)}{n^\beta}\!+\! \frac{(-1)^\beta}{ n^\beta(\beta-1)!}\times\nonumber\\
&\times\Big[\Psi^{(\beta-1)}\left(\frac{n\mp1}{n}\!+\!j\right)-\Psi^{(\beta-1)}\left(\frac{n\pm1}{n}\!+\!j\right)-\Psi^{(\beta-1)}\left(\frac{n\mp2}{n}\!+\!j\right)\Big]\Bigg\}^{-1}\nonumber\\
-&\frac{2}{\frac{2\zeta(\beta)}{n^\beta}-\frac{(-1)^\beta}{ n^\beta(\beta-1)!}[\Psi^{(\beta-1)}(\frac{n-1}{n})+\Psi^{(\beta-1)}(\frac{n+1}{n})]}.
\end{eqnarray}
If the power $\beta$ of the interaction potential is sufficiently large and if the
effect of the cut-off length $R_c$ can be disregarded, i.e. if $R_c< a/q$, the entire sum $\sum_j S^\pm_j(n)$ contributes only very little. For the case of
a van der Waals potential, i.e $\beta=6$, we have numerically verified that the sum contributes less than a few per cent to the final result. Thus
we can to a good approximation ignore the sum, which gives for the chemical potential of the $q=1/n$ phases in a system with van der Waals
interactions
\begin{eqnarray}
 \Halbnotfall\mu^{(2)}_{\pm} = \pm\, 2 n \frac{J^2}{\tilde{C}_\beta}\Bigg\{&
 -\frac{7560 n^6 (2\mp\frac{1}{n})}{16\pi^3-63[\Psi^{(5)}(\frac{n-1}{n})+\Psi^{(5)}(\frac{n+1}{n})]}\nonumber\\
&+\frac{7560 n^6(1-\frac12(\delta_{n,2}\pm\delta_{n,2}))}{8\pi^3+63[\Psi^{(5)}(\frac{n\mp1}{n})-\Psi^{(5)}(\frac{n\pm1}{n})-\Psi^{(5)}(\frac{n\mp2}{n})]}\nonumber\\
&+\frac{1}{\frac{2}{n^6}-\frac{1}{(n+1)^6}-\frac{1}{(n-1)^6}} \Bigg\},
\end{eqnarray}

\section{Exact numerical calculation}

Complementary to the perturbation analysis we perform numerical calculations for the case of a van der Waals potential. For this we employ the density-matrix renormalisation group algorithm \cite{White1992}. It is a variational technique that minimises the energy of the full Hamiltonian $\hH_0+\hH_J$ using a matrix product state (MPS) ansatz \cite{Schollwock2011}. Although limited in the amount of entanglement along the lattice that they can capture, MPS are known \cite{Verstraete2006} to approach the true ground state of one-dimensional systems quickly with growing matrix dimension, if interactions are of finite range. Although the interactions decay only polynomially in our model, we find that we can safely cut-off the interaction at a finite distance of $r$ lattice sites, and the ground state energy will be for all practical means independent of $r$ as long as we restrict the filling fractions to denominators $n$ that are small compared to $r$. In order to make our DMRG implementation more efficient we group the interaction terms as originally introduced for computations in momentum space \cite{Xiang1996}. We also take advantage of the particle number conservation, which allows us to use MPS with the proper symmetry and fix the total particle number a priory, which comes in handy to calculate the phase diagram.

\begin{figure}[t!]
\centering
a)\includegraphics[width=.8\textwidth]{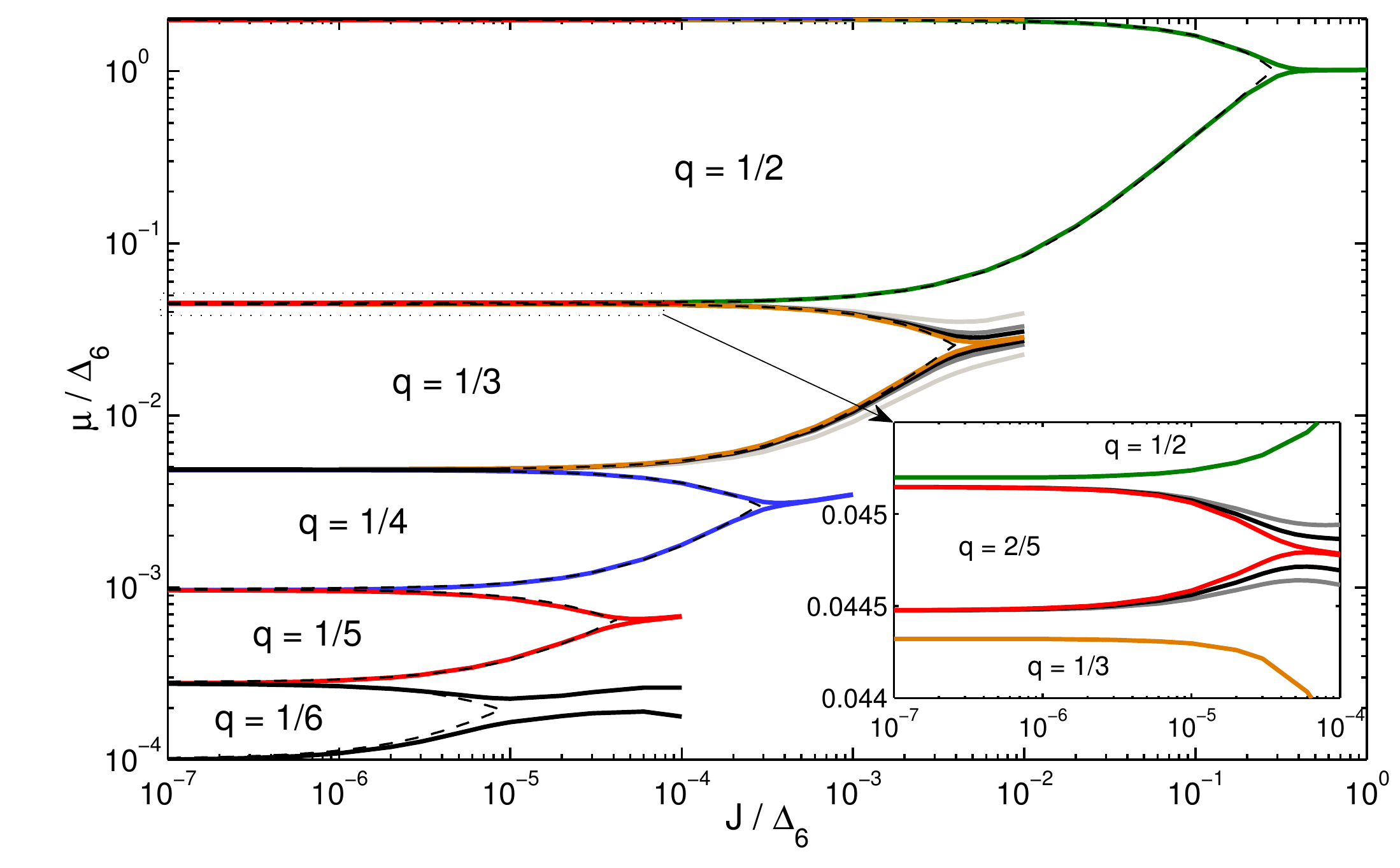}\\
b)\includegraphics[width=.8\textwidth]{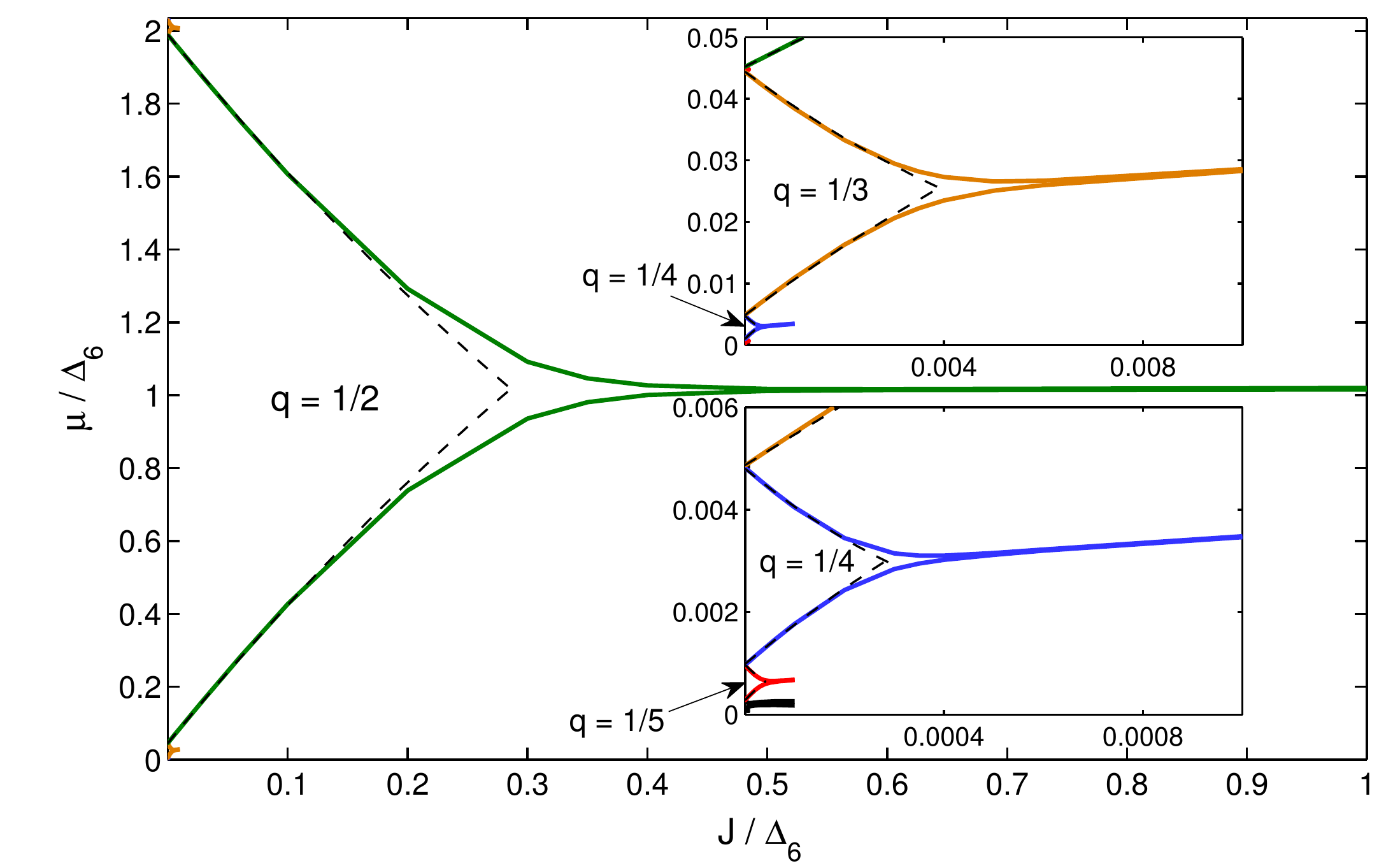}
\caption{Phase diagram for a van der Waals interaction potential. The dashed lines correspond to perturbation theory up to second order in $J/\Delta_6$, with $\Delta_6 =\tilde{C}_6/a^6$, continuous lines are calculated with the DMRG method using MPSs of dimension $32$. Coloured lines are infinite size extrapolations for $L\gtrsim24$ ($n=2,3$ only), $L\gtrsim60$, and $L\gtrsim120$ lattice sites. No infinite size extrapolation is applied for $n=6$ because we decided not to make the effort to calculate large enough system sizes. For all cases interactions over distances larger than $r=7$ lattice sites. a) Double logarithmic plot. The finite size results are shown in grey for $q=\frac13$ and $q=\frac25$ for illustration. b) Linear plot. Note the vast difference in size for different $n$ \eref{eq:mu0power}, which is much more drastic than in the $\beta=3$ case \cite{Burnell2009}.
\label{fig:phase-diagram}}
\end{figure}

To calculate the phase boundaries for a given filling fraction we make direct use of \eref{eq:mudef} by computing $E(q)$ and $E^\pm(q)$ for a finite system\footnote{We present results for open boundary conditions, i.e. no tunnelling or interaction between the left and the right end of the system, here. Periodic boundary conditions are also possible, but the lack of boundary effects does not make up for the higher numerical cost \cite{White1993} which limited our computations to smaller system sizes.}. The minimum system size $L$ to capture the physics in the thermodynamic limit for a given filling fraction $q=\frac{m}{n}$ can be estimated to be $2n^2$, because it must fit $n$ defects of size about $n$ separated from each other and the boundaries by at least one unit cell of size $n$. For not too large $n$ we can perform a infinite size extrapolation by employing different system sizes and assuming a $1/L$ scaling of the finite size error.
In \fref{fig:phase-diagram} we plot the resulting phase diagram for a van der Waals potential. The dashed lines display results from perturbation theory including up to the second order and the continuous lines are DMRG results. As expected the agreement is very good for small $J/\Delta_6$. Perturbation theory however fails close to the critical points corresponding to the tips of the lobes of incompressible phases. Also using DMRG the exact position of the 
critical point is hard to compute due to strong finite size effects. Accordingly we have not attempted to extract these values. The continuous lines in the figure end at arbitrary  values, while the phase boundary ends where these lines make contact.

As can be seen from fig.\ref{fig:phase-diagram} the melting points of the different incompressible phases
differ by many orders of magnitude in the normalized detuning $J/\Delta_6$. This raises the question if
the corresponding timescales are compatible with the finite lifetime of the dressed Rydberg atoms.
It should be noted however that the relevant time scale is determined by 
$\Delta_6 = {\widetilde C}_6/a^6$ and thus depends on the lattice constant $a$.
As mentioned in the introduction the $r^{-6}$ interaction potential is only valid beyond a cut-off length $R_c$
which sets a constraint on $a$. On the other hand the properties of the incompressible phases with filling $q$ are only determined by the tails of the interaction potential at distances $a/q$ and thus for our model
to hold it is sufficient that $a \ge R_c q$. (For the same reason the smearing out of the interaction
potential due to the convolution with the Wannier functions has little effect.)
This means $\Delta_6$ and thus the relevant time
scale can be modified by adjusting the lattice spacing according to $a\sim R_c q$ resulting in a scaling 
$\Delta_6 \sim q^{-6}$. For example taking $a= R_c q$ one finds that the melting points of the $q=1/2$ and
$q=1/6$ phases are $J_{1/2} \approx 6\times {\widetilde C}_6/R_c^6$ and $J_{1/6} \approx 0.5 \times 
{\widetilde C}_6/R_c^6$. Since ${\widetilde C}_c/R_c^6\sim \Omega^2/\Delta$, where $\Omega$ and $\Delta$
are the effective Rabi-frequencies and detunings of the Rydberg dressing, the common energy
scale of the meltig points can be tuned and can be in the kHz to MHz range. 
Thus melting can be observed also for small fillings well within the lifetime of Rydberg dressed states.

\section{Summary}

We have calculated the $\mu-J$ phase diagram for Rydberg dressed atoms in a deep one-dimensional lattice potential. For vanishing
hopping strength, $J=0$, the chemical potential $\mu$ forms a complete devils staircase as function of the filling fraction
corresponding to stable crystalline phases at any rational filling. 
Following a similar analysis of reference \cite{Bak1982} we derived a closed analytic expression for any power-law interaction potential.
We then analysed the melting of the Rydberg crystals with increasing hopping. To this end we performed a second order strong-coupling expansion
and found excellent agreement to exact numerical simulations based on DMRG adapted to long-range interactions. For the case of
stronger localised interactions, such as a van der Waals coupling, a compact analytic expression for the phase boundaries of
phases with filling $q=1/n$ was found.

\ack

We thank J. Otterbach for inspiring discussions. This work was supported by the SFB TR49 of the Deutsche Forschungsgemeinschaft and the Graduate School of Excellence MAINZ.

\section*{References}


\end{document}